\documentclass[11pt]{article}
\usepackage{hyperref}
\usepackage{authblk}
\usepackage{graphicx}
\usepackage{caption}
\usepackage{amsmath}
\usepackage{amssymb}
\usepackage{mathtools}
\usepackage{multirow}
\usepackage{amsmath}
\usepackage{color}
\usepackage{subcaption}
\usepackage{wasysym}
\usepackage{ulem}

\begin{document}

\title{Reaction of deuteron with a heavy nucleus at low energies}

\author{Pankaj Jain$^{1}$\footnote{pkjain@iitk.ac.in}}
\author{Harishyam Kumar$^{2}$\footnote{hari@iitk.ac.in}}
\author{K. Ramkumar$^{2}$\footnote{kaanapuliramkumar08@gmail.com}}
\affil{
	$^1$ Department of Space, Planetary and Astronomical Sciences \& Engineering, Indian Institute of Technology, Kanpur, 208016, India\\
$^2$ Physics Department, Indian Institute of Technology, Kanpur, 208016, India
}

\maketitle

\begin{abstract}
 We extend the recently proposed mechanism for inducing low energy nuclear reactions (LENR) to compute the reaction rate of deuteron with a heavy nucleus. The process gets dominant contribution at second order in the time dependent perturbation theory and is assisted by a resonance. The reaction proceeds by breakdown of deuteron into a proton and a neutron due to the action of the first perturbation. In the second, nuclear perturbation, the neutron gets captured by the heavy nucleus. Both perturbations are assumed to be electromagnetic and lead to 
emission of two photons, one at each vertex. The heavy nucleus is taken to
	be ${}^{58}$Ni although many other may be considered.
	The reaction rate is found to be very small
unless assisted by some special conditions. In the present case we assume 
the presence of a nuclear resonant state. In the presence of such a state
we find that the reaction rate 
 is sufficiently large  to be observable in laboratory even at low energies.
\end{abstract}

\section{Introduction}

Nuclear fusion are expected to be strongly suppressed at low energies
due to the Coulomb barrier \cite{1968psen.book.....C}. However, there
exists 
 considerable experimental evidence that such reactions may occur at observable
 rates even at very low energies
 \cite{PhysRevC.78.015803,StormsCS2015,McKubre16,Cellani19,Mizuno19,SRINIVASAN2020233}. The current status is reviewed in \cite{doi:10.1002/9781118043493.ch41,doi:10.1002/9781118043493.ch42,doi:10.1002/9781118043493.ch43,biberian2020cold}. 
Theoretically, there have been many attempts to explain these processes in terms of electron screening \cite{PhysRevC.101.044609,assenbaum1987effects,ichimaru1993nuclear}, correlated states \cite{articleVy,PhysRevAccelBeams.22.054503}, electroweak interactions \cite{Widom10}, formation of clusters of nuclear particles \cite{SPITALERI2016275}, relativistic electrons in deep orbits \cite{Meulenberg19} and phonon induced reactions \cite{Hagelstein15}.
A critical review of many claims in this field is provided in \cite{Chechin_1994}.

It has been proposed that there exist additional processes which open up at
 second order in the perturbation theory for which the rate may be significant
even at low energies. 
\cite{PhysRevC.99.054620,Jain2020, Jain2021, Ramkumar2022}. These reactions
are different from the standard first order fusion process which is heavily
suppressed \cite{1968psen.book.....C}. 
The basic idea is that the reaction proceeds due to two interactions 
with widely different distance scales. The first perturbation causes the system to go into a state
which is a linear superposition of all eigenstates of the unperturbed Hamiltonian. Due to the presence of eigenstates of relatively high energy, it is possible that the Coulomb barrier may not be a very serious issue. 
Although the amplitude for such high energy eigenstates is suppressed, 
the suppression may not be as strong as that due to the
Coulomb barrier. The resulting amplitude gets contribution from relatively
large atomic scale distances and we refer to it as the molecular matrix 
element. The second perturbation leads to the nuclear transition and gets
contribution from small distances. We refer to it as the nuclear matrix 
element.

 We applied this formalism explicitly to the process involving fusion of proton with deuteron to form helium nucleus with $A=3$ \cite{Jain2020,Jain2021}. 
The perturbation was assumed to be electromagnetic leading to either emission or absorption of photons. The dominant process was found to be the one in which two photons are spontaneously emitted. The rate was found to be very small in free space due to cancellation among different eigenstates contributing to
the intermediate
state. Similar phenomenon is seen for other low energy nuclear processes
\cite{Kumar:2023qdb,ramkumar2024low,jain2024medium}.
However it was argued that in a medium, under special conditions, the rate may be significant and observable \cite{Jain2020,Jain2021,Ramkumar2022,Kumar:2023qdb,jain2024medium}.
This mechanism has also been used to provide the theoretical explanation for
the experimentally observed low energy reaction of Palladium with Deuteron 
in medium to
produce Silver and neutron \cite{gadly2024transmutation}. 

It has recently been shown that the rate may
be substantial, even in free space, in the presence of a resonance \cite{ramkumar2024low}. In this paper we considered the reaction in which an incident 
proton converts into a neutron through weak interactions and the neutron 
gets absorbed by a heavy nucleus emitting a photon. We applied this explicitly
to the case of nickel nucleus with A=58 which has a resonance at 7 keV with 
width of approximately 7 eV. The rate for this process was found to be 
relatively large and observable.
Here we extend this mechanism to another process involving fusion of deuteron ${}^2$H and a heavy nucleus ${}^A$X of atomic number $Z$ and atomic mass $A$. The process we shall consider is the following:
\begin{equation}
	^{2}{\rm H} +  {}^{A}{\rm X} \rightarrow {}^{A+1}{\rm X} + {}^1{\rm H} + Y 
    \label{eq:HXfusion}
\end{equation}
where $Y$ may consist of a wide range of final states. 
In this paper we shall confine ourselves to a particular process in which
the final state involves two photons. 
 In this case,
\begin{equation}
Y = \gamma(\omega_1) + \gamma(\omega_2)\,.
\label{eq:HXfusionY}
\end{equation}
and both perturbations are electromagnetic.

\begin{figure}
     \centering
     \includegraphics[width=0.84\textwidth]{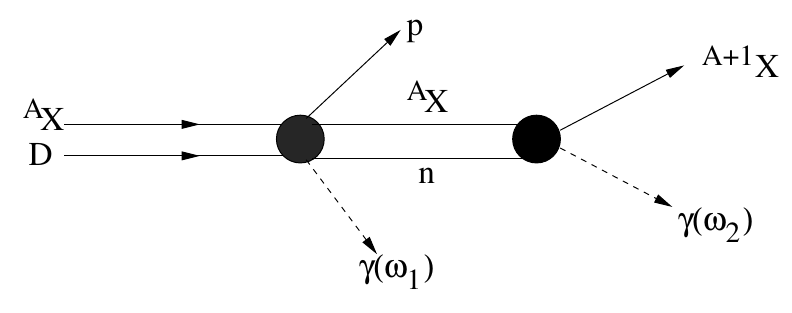}\\
     \caption{Schematic illustration of the reaction described in Eq. \ref{eq:HXfusion}. 
}
\label{fig:process}
\end{figure}

Within the framework of the second order perturbation theory, the 
reaction proceeds in two steps. Due to the action of the first perturbation, the deuteron forms an intermediate state with emission of a photon and a proton. 
We refer to this as the first vertex, as shown in Fig. \ref{fig:process}. The second perturbation leads to capture of neutron by the nucleus ${}^A$X with emission of another photon. We refer to this as the second vertex. 
We point out that the photon with frequency $\omega_1$ can also be emitted from the second 
vertex and that with frequency $\omega_2$ from the first vertex. 
The intermediate
state, formed after the action of the first perturbation, involves 
superposition of all eigenstates of the unperturbed Hamiltonian. 
The $^2$H is initially in the ground state. The next state available is the one in which the neutron and the proton are barely free. This state corresponds to zero kinetic energies for both of these particles. In the second order perturbation theory we need to sum over all intermediate eigenstates, 
 up to infinite energy.

\section{Deuteron-X fusion at second order in perturbation theory}

The Hamiltonian of the system is given by 
\begin{equation}
    H  = H_0 + H_I
\end{equation}
where $H_0$ denote the 
unperturbed Hamiltonian and $H_I$ is a time dependent perturbation. 
The unperturbed part contains the kinetic energy terms and screened Coulomb and nuclear potentials. 
We assume a simple spherical well model for the neutron-proton nuclear potential. Hence
for small separations, the potential experienced by the neutron-proton 
system is given by
\begin{equation}
	V_{pn} = -V_0
	\label{eq:potentialD}
\end{equation}
for $r<r_0$ and $V_{pn}=0$ for larger distances. Here we set $V_0=34.6$ MeV
and $r_0=2.066$ fm which leads to the deuteron bound state of correct
binding energy. For the nuclear potential of the $X$ nucleus we use the
shell model potential, same as that used in \cite{ramkumar2024low},
\begin{equation}
	V_{X} = -{V_{0}\over 1+\exp[(r-R_X)/a_X]}
	\label{eq:potentialX}
\end{equation}
where, $V_{0} = 50$ MeV, $a_X = 0.524$ fm and $R_X=1.25A^{1/3}$ \cite{Krane:359790}. This leads to the nickel 2p state at energy of -9.2 MeV, in good 
agreement with the observed value of -9 MeV for ${}^{59}$Ni.

We denote the coordinates of proton, neutron and the X nucleus by 
$\vec r_1$, $\vec r_2$ and $\vec r_X$ respectively. We define the
relative coordinate $\vec r = \vec r_2 - \vec r_1$ and denote the
center of mass coordinate of the proton-neutron system by $\vec R_{cm}$. 
The overall center of mass coordinate of the three particle system is
denoted by $R'_{cm}$. 
The initial state wave function can be expressed as
\begin{equation}
	\Psi_{i} = \Psi_{cm}(\vec R'_{cm}) \psi_D(\vec r,\vec R_{cm})
\end{equation} 
where $\Psi_{cm}(\vec R'_{cm})$ is the overall center of mass wave function
and $\psi_D$ is the deuteron wave function.
We assume that in the initial state the deuteron and X atoms
form a molecular bound state. 
We may express the deuteron wave function as
\begin{equation}
	\psi_D(\vec R_{cm},\vec r) = \psi_{i}(\vec R_{cm}) \chi_D(r)
	 = \psi_{i}(\vec R_{cm})  {u_D(r)\over \sqrt{4\pi}\, r}
\end{equation}
where $\chi_D(r)$ is the deuteron nuclear ground state 
wave function corresponding to $l=0$ and 
$\psi_{i}$ is the molecular bound state wave function of the deuteron-X 
system.
We shall assume that the molecular wave function $\psi_{i}$ is spherically symmetric. The nucleus $X$ is assumed to be very heavy and we ignore its
recoil, which is expected to be a good approximation. 
We point out that the overall center of mass motion does not
add any essential detail to the calculation and only impose overall momentum 
conservation. 

The interaction Hamiltonian 
 \cite{merzbacher1998quantum,sakurai1967advanced} is given by
\begin{equation}
    H_I(t) = 
\sum_i{ Z_ie\over c m_i} \vec A(\vec r_i,t)\cdot \vec p_i  
    \label{eq:Hint}
\end{equation}
where $Z_i$, $m_i$, $\vec r_i$ and $\vec p_i$ are respectively the charge, mass, position vector and momentum vector of the particle $i$. 
The electromagnetic field operator $\vec A(\vec r,t) $ is given by
\begin{equation}
    \vec A(\vec r,t) = {1\over \sqrt{V}} \sum_{\vec k}\sum_\beta c\sqrt{\hbar\over 2\omega} \left[a_{\vec k,\beta}(t) \vec\epsilon_\beta e^{i\vec k\cdot\vec r} +  a^\dagger_{\vec k,\beta}(t) \vec\epsilon^{\,*}_\beta e^{-i\vec k\cdot\vec r}\right]
\end{equation}
%\begin{figure}
%     \centering
%     \includegraphics[width=0.84\textwidth]{coord1}\\
%     \caption{The nucleus X is centered at the origin with $\vec R_{cm}$ and $\vec r$ being the center of mass and relative coordinates of deuteron. }
%\label{fig:coord1}
%\end{figure}
The leading order contribution to the process in Eq. \ref{eq:HXfusion} with $Y$ given by Eq. \ref{eq:HXfusionY}
is obtained at second order in the time dependent perturbation theory. Let the wave vectors of the two emitted photons be $\vec k_1$ and $\vec k_2$ and their frequencies $\omega_1$ and $\omega_2$ respectively. The
transition amplitude at this order 
can be expressed as,
\begin{equation}
	\langle f|T(t_0,t)|i\rangle = \left( -{i\over \hbar}\right)^2\sum_{E_n} \int_{t_0}^t dt' e^{i(E_f-E_n)t'/\hbar} \langle f|H_I(t')|n\rangle \int_{t_0}^{t'} dt'' e^{i(E_{p}+E_n-E_i)t''/\hbar} \langle pn|H_I(t'')|i\rangle
    \label{eq:TransitionII} \ .
\end{equation}
where the sum is over the intermediate neutron states.
Here $|pn\rangle$ denotes the intermediate state proton and neutron 
wave function with 
 $E_p$, $E_n$ being the energies of the proton and neutron states
 respectively produced
at the first vertex. 

\subsection{Molecular Matrix Element}
Let us first consider the transition from initial to intermediate state
which we refer to as the molecular matrix element.
We note that in the sum over $i$ in Eq. \ref{eq:Hint} only the proton contributes while computing the matrix element in the integral over $t''$. Either of the two photons can emerge from this interaction leading to two amplitudes. Let us first write to the amplitude corresponding to emission of photon with frequency $\omega_1$. 
The $t''$ integral then gives
\begin{eqnarray}
	\int_{t_0}^{t'} dt'' e^{i(E_p + E_n-E_i)t''/\hbar} \langle pn|H_I(t'')|i\rangle &=& -{ie\hbar\over m_p }\sqrt{\hbar\over 2\omega_1 V} \langle pn|\vec\epsilon^{\,*}_\beta\cdot \vec p_1\ |i\rangle \nonumber\\ &\times& {e^{i(E_p+E_n-E_i+\hbar\omega_1)t'/\hbar} \over E_p+E_n-E_i+\hbar \omega_1 }
\label{eq:inttpp}
\end{eqnarray}
where we have dropped the term depending on the arbitrary initial time $t_0$ which is expected to give negligible contribution \cite{merzbacher1998quantum,sakurai1967advanced}. 
The matrix element can be written as \cite{sakurai1967advanced}
\begin{equation}
\langle pn|\vec\epsilon^{\,*}_\beta\cdot \vec p_1\ |i\rangle = 
	{im_1\over \hbar}(E_p + E_n-E_i)\,
\langle pn| \vec\epsilon^{\,*}_\beta\cdot\vec r_1\ |i\rangle
\label{eq:matrixni}
\end{equation}
We next express $\vec r_1$ as
\begin{equation}
 \vec r_1 = \vec R_{cm} -{m_2\over M}\ \vec r
\end{equation}
where $M=m_1+m_2$, with $m_1=m_p$ and $m_2=m_n$. 
Let us assume that the $R_{cm}$ 
 dependence of the wave function is given by,
\begin{equation}
	\psi_{i} = {N_{cm}\over \sqrt{4\pi}} \ 
	e^{-(R_{cm} - R_0)^2/\Delta^2}
 \label{eq:psicm}
\end{equation}
where $N_{cm}$ is the normalization factor. 
This is essentially the molecular bound state wave function of the 
D-X system and is controlled by the molecular potential. Here we directly
assume the form of the wave function, applicable at 
 $R_{cm}$ larger than roughly 1 atomic units. 
For smaller $R_{cm}$ values the form will be modified by the Coulomb 
repulsion and at nuclear distances by the nuclear potential. However, 
this region contributes negligibly to the molecular matrix element and 
hence the difference can be ignored.
The assumed form is convenient
since it allows analytic calculations. A more realistic form is expected to
have a similar behavior and hence will not lead to substantially different 
results.
We consider values of 
$R_0$ and $\Delta$ roughly equal to 3.5 and 0.3 in atomic units, which 
are reasonable for a molecular bound state. Note that with these values
the wave function is heavily suppressed for distances smaller than about 1 
atomic unit.

The matrix element on the right hand side of Eq. \ref{eq:matrixni} can now be computed by solving the Schrodinger equation to obtain all eigenstates of the proton neutron system in terms of the relative coordinate $\vec r$. 
Here the initial state is the ground state, i.e. the deuteron. 
We ignore the $l=2$ component of deuteron and set $l=0$, where $l$ is the orbital quantum number. Hence the initial state has quantum numbers $S=1$, $l=0$ and $j=1$.   
The dominant contribution from the intermediate state is obtained with orbital quantum number $l=1$.  
Hence, with $S=1$ we can have $j=0,1,2$. 

Let us consider the initial state with $m=1$ where 
$\langle j_z\rangle = \hbar m$. The spin part of the wave function
is not affected in the entire process. Hence both the proton and neutron
have spin up for the entire process and here we focus entirely on the 
space part of the wave function. 
Let us choose the photon momentum $\vec k_1$ such that,
\begin{equation}
	\vec k_{\gamma 1} = k_{\gamma 1} \left(\cos\theta_{\gamma 1} \hat z +\sin\theta_{\gamma 1}(\cos\phi_{\gamma 1}
	\hat x + \sin\phi_{\gamma 1}\hat y)\right] 
\end{equation}
The two polarization vectors of this photon can be expressed as
\begin{eqnarray}
	\vec\epsilon_{1a} &=& 
	 -\sin\theta_{\gamma 1} \hat z +\cos\theta_{\gamma 1}
	 (\cos\phi_{\gamma 1}
	\hat x + \sin\phi_{\gamma 1}\hat y) \nonumber\\
	\vec\epsilon_{1b} &=& -\sin\phi_{\gamma 1}\hat x + \cos\phi_{\gamma 1}\hat y
\end{eqnarray}
 Here we consider only the polarization vector
 $\vec \epsilon_{1a}$. The other polarization will add incoherently and can
 only enhance the computed result.
We can express the matrix element as
\begin{equation}
\langle pn| \vec\epsilon^{\,*}_\beta\cdot\vec r_1 |i\rangle = 
	\int \int d^3 r d^3R_{cm} \Psi^*_{pn}(\vec r, \vec R)\vec\epsilon^{\,*}_{1a}
	\cdot\vec r_1 e^{-i\vec k_{\gamma 1}\cdot \vec r_1}\psi_{i}(R_{cm})\chi(r)
\end{equation}
Here $\Psi_{pn}$ is the intermediate state wave function of the proton-neutron
system. It is a wave function
of neutron and proton in the potential due to the $X$ particle. 

The wave function $\Psi_{pn}$ is
somewhat complicated and we make some simplifying assumptions. 
Let us first consider this wave function in the absence of the particle $X$. 
In this case the $\vec R_{cm}$ dependence is expected to be a plane wave
and we can express it as,
\begin{equation}
	\Psi_{pn} ={1\over \sqrt{V}} 
	e^{i\vec k_+\cdot \vec R_{cm}}\chi_{pn}(\vec r) 
	\label{eq:Psimsmallr}
\end{equation}
where $\vec k_+ = \vec k_1 + \vec k_2$ and $\vec k_1$ and $\vec k_2$ are the proton and 
neutron wave vectors respectively. 
The wave function $\chi_{pn}$ is the  
$l=1$ proton-neutron wave function in the nuclear potential corresponding to 
$E>0$. It depends on $\vec k_-=(\vec k_2-\vec k_1)/2$ and 
we express it as,
\begin{equation}
	\chi_{pn} = {u_{pn}(r)\over r} \zeta(\hat r)
	\label{eq:Psimsmallr1}
\end{equation}
where the angular dependence is contained in $\zeta(\hat r)$ and can be read off
from $l=1$ part of the plane wave expansion
\begin{equation}
	e^{i\vec k_-\cdot \vec r} = 4\pi \sum_{l=0}^{\infty}
	\sum_{m=-l}^{l} i^l j_l(k_-r)\ Y_l^{m}(\hat r)
	Y_l^{m*}(\hat k_-)
	\label{eq:PlaneWave}
\end{equation}
We point out that we are not explicitly displaying
the spin part of the wave function since it does not play any role in our
calculation. 
The wave function $\Psi_{pn}$ gets modified due to the presence of the
particle X and the factorized form assumed in Eq. \ref{eq:Psimsmallr} is
not applicable.
 We are interested in the final state in which proton is free
and neutron gets captured by X. Hence a reasonable approximation is 
to assume a plane wave form for the proton and determine the neutron 
wave function by using Schrodinger equation. For the molecular matrix
element we require this wave function only 
at large distances from X, i.e. large $R_{cm}$. 
 This is because the 
initial state wave function is strongly suppressed
at small $ R_{cm}$. Furthermore, the dominant contribution is obtained
for very small values of the relative coordinate $r=|\vec r|$. Hence it is 
reasonable to assume following form of the wave function,
\begin{equation}
	\Psi_{pn} ={1\over \sqrt{V}} \psi_n(R_{cm}) 
	e^{i\vec k_1\cdot \vec R_{cm}}\chi_{pn}(\vec r) 
	\label{eq:Psimsmallr2}
\end{equation}
Here, in the intermediate state $\psi_n$ refers to the neutron wave function
in the presence of the particle X. 
It depends on the variable $\vec r_2$. However
since only very small values of $r$ contribute to molecular matrix element,
it is reasonable to assume $\vec r_2\approx \vec R_{cm}$. Similarly we have 
assumed $\vec r_1\approx \vec R_{cm}$.
The wave function $\psi_n$ may be obtained by solving the Schrodinger equation
for neutron in the presence of particle X.
As we shall have seen in \cite{ramkumar2024low}, a significant contribution is obtained
only in the presence of a resonance in the nuclear matrix element. This is applicable 
as long as we assume free space wave functions at large distances. In order to 
include the contribution of resonance, here we directly model the phase shift in the 
wave function \cite{ramkumar2024low}, as explained later in this section.

The matrix element can be written as
\begin{equation}
\langle pn| \vec\epsilon^{\,*}_\beta\cdot\vec r_1\ |i\rangle
= 
\langle pn| \vec\epsilon^{\,*}_\beta\cdot\vec R_{cm}\ |i\rangle - {m_2\over M} 
	\langle pn| \vec\epsilon^{\,*}_\beta\cdot\vec r\ |i\rangle
	\label{eq:decompose}
\end{equation}
Let us first consider the matrix element corresponding to $\vec R_{cm}$. 
We have 
\begin{equation}
	\langle pn| \vec\epsilon^{\,*}_\beta\cdot\vec R_{cm} |i\rangle = 
	\int \int d^3 r d^3R_{cm} \Psi^*_{pn}(\vec r, \vec R)\vec\epsilon^{\,*}_{1a}
	\cdot\vec R_{cm} e^{-i\vec k_{\gamma 1}\cdot \vec R_{cm}} 
	 e^{i\vec k_{\gamma 1}\cdot \vec rm_2/M} 
	\psi_{i}(R_{cm})\chi_D(r)
\end{equation}
We get dominant contributions only from very small values of $r$, of
the order of the deuteron radius. Hence 
the exponential factor involving $r$ can be set to unity at leading order. 
The integral over $r$ then involves a direct overlap between the deuteron
bound state wave function and a free proton-neutron wave function corresponding
to deuteron potential. This clearly vanishes and hence, to leading order,
this matrix element is zero. We therefore focus on the second
matrix element on the right hand side of Eq. \ref{eq:decompose} which involves
$\vec r$. This matrix element can be written as,
\begin{eqnarray}
	\langle pn| \vec\epsilon^{\,*}_\beta\cdot\vec r |i\rangle &=& {1\over \sqrt{V}} 
	\int \int d^3 r d^3R_{cm} \chi_{pn}^{*}(\vec r) 
	e^{-i\vec k_+\cdot \vec R_{cm}}\vec\epsilon^{\,*}_{1a}
	\cdot\vec r e^{-i\vec k_{\gamma 1}\cdot \vec R_{cm}} 
	 e^{i\vec k_{\gamma 1}\cdot \vec rm_2/M} 
	\psi_{i}(R_{cm})\chi_D(r)\nonumber\\
	&=& I_a I_b
\end{eqnarray}
where 
\begin{equation}
	I_a =  	\int d^3 r \chi_{pn}^{*}(\vec r) \vec\epsilon^{\,*}_{1a}
	\cdot\vec r  e^{i\vec k_{\gamma 1}\cdot \vec rm_2/M} 
	\chi_D(r)
\end{equation}
and
\begin{equation}
	I_b =  {1\over \sqrt{V}} 
	 \int d^3R_{cm} \psi^*_n(R_{cm})  
	e^{-i\vec k_1\cdot \vec R_{cm}}  
	 e^{-i\vec k_{\gamma 1}\cdot \vec R_{cm}} 
	\psi_{i}(R_{cm})
\end{equation}

The wave function $\chi_{pn}$ depends on the wave vector $\vec k_-$ and is 
normalized to a plane wave. We denote
the angular coordinates of $\vec k_-$ as $(\theta_-,\phi_-)$. We get contributions
from all the three components $l_z=-1,0,1$. 
Performing the angular integrals in $I_a$, we find that 
\begin{equation}
	I_a = {1\over 2}I_\Omega I'_a
	\label{eq:Ia1}
\end{equation}
where 
\begin{equation}
I'_a = \int dr u^*_{pn}(r) r u_D(r)
\end{equation}
and
\begin{equation}
	I_{\Omega} = -\cos\theta_- \sin\theta_{\gamma 1} + 
	\sin\theta_-\cos\theta_{\gamma 1} \cos(\phi_--\phi_{\gamma 1}) 
\end{equation}
The radial integral in Eq. \ref{eq:Ia1} gets dominant
contribution from the leading order term proportional to $j_1$
and is approximately proportional $k_-$ at low energies relevant 
to our calculation.  
The integral $I_b$ can be expressed as,
\begin{equation}
	I_b =  {4\pi\over \sqrt{V} K}  \int_0^\infty dR_{cm}  
	u^*_n(R_{cm})\sin(KR_{cm})e^{-(R_{cm}-R_0^2)^2/\Delta^2}
\end{equation}
where $K=|\vec k_1+\vec k_{\gamma 1}|$ and we have expressed $\psi_n=u_n/R_{cm}$. 

At large distances, the intermediate state neutron wave function can be written as
\begin{equation}
	u_n(r_2) = \sin(k_2 r_2 + \delta(k_2)) 
\end{equation}
where $\delta(k_2)$ is the phase shift. Here we directly model the phase shift
for the neutron-nickel system by taking
guidance from the observed data. This system shows the first resonance at 7 keV which 
has a width of about 7 eV. We denote the value of $k_n$ corresponding to the resonance
as $k_R$. We are interested in a phase shift which accurately models this
resonance. The standard expansion of the phase shift for $l=0$ can be written as,
\begin{equation}
k\cot\delta = -a_1 + a_2 k^2 + ...
\label{eq:reso}
\end{equation}
We use the expansion up to $k^6$ and model the phase shift as,
\begin{eqnarray}
	\sin\delta  &=& {-k_n\over  \sqrt{k_n^2 + (a_1-a_2 k_n^2)^2
	(1+b_1k_n^2+b_2k_n^4)^2}}\nonumber\\
	\cos{\delta} &=& {(a_1-a_2k_n^2)(1+b_1k_n^2+b_2k_n^4)\over \sqrt{k_n^2 + (a_1-a_2 k_n^2)^2(1+b_1k_n^2+b_2k_n^4)^2}}
\label{eq:resodelta}
\end{eqnarray}
Here we have isolated the factor 
$a_1-a_2k_n^2$ which leads to a resonance at $k_n= k_R=\sqrt{a_1/a_2}$. 
This is a generalization of the model used in \cite{ramkumar2024low} and allows us to
properly model the height of the resonance. 
As expected, $\delta=0$ in the limit $k_n$ goes to zero. 
We can now directly choose the parameters $a_1$, $a_2$, $b_1$ and $b_2$ so that
the phase shift is properly modelled at low energies, as explained later.  
The main assumption in this model is that the $k_n$ dependence of the nuclear
matrix element is dominated by the resonant term which is given by the 
denominator in Eq. \ref{eq:resodelta}.

Here, as well as \cite{ramkumar2024low}, we are directly modelling the resonance
in terms of the phase shift and not using a 
potential model. We may also consider some simple potential models to determine if they lead to a  
result which is not negligible. Here we consider a square well potential, with $V=-V_0$ for $r<r_0$ and
$V=0$ for $r\ge r_0$. In general this potential produces resonances with rather wide width which will lead to an
amplitude close to zero. However, it is possible to find a resonance close to $E=0$, with energy slightly
larger than zero and with a relatively narrow width. Such
a resonance exists, for example, for parameters, $V_0=30 $ MeV and $r_0 = 2.19529\times 10^{-4}$ atomic units (approximately
10 fm). The nuclear bound state energy for these parameters is found to be $-22.2$ MeV. 
We find many such parameter choices which allow presence of a low energy 
resonance. For such cases, we find a non-negligible contribution for small $K$ of order unity. 
In our calculation we did not include the neutron bound states in the
sum over intermediate states. However, these states are highly suppressed
at large distances and will lead to extremely small molecular matrix elements.
Hence, it is reasonable to neglect the contribution from such states.
A more detailed investigation
of potential models is postponed to future research.

\subsection{Nuclear Matrix Element}
We next consider the nuclear matrix element $\langle f|H_I(t')|n\rangle$. 
In this case, the intermediate state neutron undergoes 
fusion with the heavy nucleus. As for the case of free neutron, a wide range of processes associated with standard neutron capture can take place. Here we shall focus on the process with photon emission. The process is complicated since as the neutron interacts with the nucleus, it can interact with all the nucleons, leading to a change in the multiparticle wave function. Furthermore, the photon can be emitted by any one of the proton or neutron in the system. 
Here we assume that we can treat the nucleus $X$ as a single particle of charge
$Z_X$. 
 We take the emitted photon momentum to be $\vec k_{\gamma 2}$. 
Let the angular coordinates of this photon momentum be $(\theta_{\gamma 2}, \phi_{\gamma 2})$.  
We point
out that we also need to add the amplitude for which photon one and two
are interchanged, i.e. photons with frequencies $\omega_1$ and
$\omega_{ 2}$ are emitted from second and first vertex in Fig. \ref{fig:process} respectively. 

In order to proceed further we need to specify the final state eigenfunction.
We take this state to be $l=1$, $j=3/2$ shell model 
state with one unpaired neutron in the
outer most shell. As stated above we shall consider the state for which 
neutron has spin up and set $j_z=3/2$. 
We point out that ${}^{58}$Ni (spin 0)
satisfies
our requirements. It has 30 neutrons which means that it has two neutrons 
in the outermost $2p_{3/2}$ level. Adding one more neutron in the $l=1$ 
level will lead to a total spin $3/2$ with $j_z=3/2$. 
Using the square well potential (Eq. \ref{eq:potentialX}), we find that the
energy eigenvalue of this state is 9.1 MeV in good agreement with the observed value.
We point out that
we can get transitions into other states also but this choice is convenient,
as explained below. Furthermore the heavy nucleus need not be Ni. It would
be interesting to consider other possibilities.

We perform the calculation by specializing to a particular polarization
vector of the emitted photon with frequency $\omega_2$. This is taken to
be 
\begin{equation}
\vec\epsilon_{2b} = -\sin\phi_{\gamma 2}\hat x + \cos\phi_{\gamma 2}\hat y
\end{equation}
The photon is emitted only by the particle $X$. We denote its position vector
by $\vec r_X$. Let $\vec R'_{cm}$ and $\vec r'$ be respectively
the center of mass and relative coordinate of the neutron-$X$ system.  
The matrix element can be expressed as,
\begin{equation}
	\langle f|H_I(t')|n\rangle = {iZ_Xe\over \hbar}\sqrt{\hbar\over
	2V\omega_2}e^{iE_{\gamma 2}t'/\bar h} (E_f-E_{n})
	\langle f|\vec \epsilon^*_{2b}\cdot \vec r_X|n\rangle
\end{equation}
where $|n\rangle$ represents the neutron intermediate state of energy 
$E_n$. We express $\vec r_X$ in terms of the center of mass coordinate
$\vec R'_{cm}$ and relative coordinate $\vec r'$ of the neutron-$X$ system,
\begin{equation}
	\vec r_X = \vec R'_{cm} - {m_2\over M'} \vec r'
\end{equation}
where $M'=m_X+m_2$ is the mass of the final state nucleus. 

We first consider
\begin{equation}
\langle f|\vec \epsilon^*_{2b}\cdot \vec R'_{cm} |n\rangle
	= \int d^3R'_{cm} d^3r' \psi^*_f(\vec r')\psi^*_{fc}(\vec R') \vec\epsilon^*_{2b}
	\cdot \vec R'_{cm} e^{-ik_{\gamma 2}[\vec R'_{cm}-{m_n\over M'}\vec r']}  \psi_{nX}(\vec r')\psi_{nXc}(\vec R')
\end{equation}
Here $\psi_{nX}$ and $\psi_{nXc}$ are the intermediate state wave functions of the 
neutron-$X$. Similarly $\psi_f$ and $\psi_{fc}$ are the final state wave functions.  
Let us consider the integral over $r'$. Since the final state wave function
is significant only for very small values of $r'$, we can drop the $\vec r'$ dependent
term in the exponential to leading order. The integral then vanishes since the
two $\vec r'$ dependent wave functions are orthogonal to one another. 
We next consider the $\vec r'$ dependent term. We obtain
\begin{equation}
\langle f|\vec \epsilon^*_{2b}\cdot \vec r' |n\rangle
	= \int d^3R'_{cm} d^3r' \psi^*_f(\vec r')\psi^*_{fc}(\vec R') \vec\epsilon^*_{2b}
	\cdot \vec r' e^{-ik_{\gamma 2}[\vec R'_{cm}-{m_n\over M'}\vec r']}  \psi_{nX}(\vec r')\psi_{nXc}(\vec R')
\end{equation}
Integration over the center of mass variable just imposes the overall momentum conservation and we focus on the relative variable. 
Again we can drop the $\vec r'$ dependence in the exponential function.
The state $\psi_f(\vec r')$ is the $l=1$ state of the shell model nuclear
wave function, as discussed above. The state $\psi_{nX}$ is taken to the
$l=0$ state which gives dominant contribution for energies under consideration. 
Note that the neutron-proton system has $l=1$. The neutron intermediate state can
take all $l$ values which can lead to $l=1$ for the neutron-proton system. However the
values $l\ne 0$ for the neutron state lead to very small amplitudes and can be neglected.
The matrix element can be written as,
\begin{equation}
\langle f|\vec \epsilon^*_{2b}\cdot \vec r' |n\rangle
	= I'_{\Omega} I_f 
\end{equation}
where
\begin{equation}
I_f	= \int dr' u^*_f(r')
	r'  u_{n}(r')
\end{equation}
and
\begin{equation}
	I'_{\Omega} = i\, \sqrt{2\pi\over 3}e^{-i\phi_{\gamma 2}}
	\end{equation}
Here we have expressed $\psi_f = Y^1_1(\hat r')u_f(r')/r'$ and
$\psi_{nX} = Y^0_0 u_{n}(r')/r'$. 

We estimate the nuclear matrix element $I_f$ at $k_n=0$ using the shell model
potential for the wave functions given in Eq. \ref{eq:potentialX}. However to determine
its dependence on $k_n$ we directly use the model for the phase shift given in 
Eq. \ref{eq:resodelta}. We note that $I_f$ is expected to be almost independent 
of $k_n$ at low energies. At medium energies it has a mild dependence and shows
dramatic change close to resonance. This is captured by the following
overall factor in the wave function $u_n$ at very small $r'$,
corresponding to nuclear distances,
\begin{equation}
	{a_1 \over \sqrt{k_n^2+(a_1-a_2k_n^2)^2(1+b_1k_n^2+b_2k_n^4)^2}}
\end{equation}
which goes to unity in the limit $k_n\rightarrow 0$. 
This will lead to the proper behaviour of the standard
neutron absorption cross section cross section both at low energy and in
the vicinity of the resonance.
The parameters are taken to be $a_1=1.6\times 10^4$, $a_2= 1.2\times 10^{-2}$, 
$b_1 = 0$ and
$b_2= 8.0\times 10^{-11}$, which correctly model the observed resonance
in neutron-nickel system and energy of 7 keV. Here we have set $b_1=0$ for
simplicity. This parameter can be adjusted by making a more detailed fit
to the observed neutron absorption cross section. Here we are only interested in the contribution
from resonance and it is reasonable to set it to zero.
We point out that the estimated value of $I_f$, using the shell model potential,
leads to a good agreement of the standard neutron absorption cross section on
${}^{58}$Ni with photon emission,
which is found to be about 200 barns at incident energy of
$10^{-5}$ eV.

\section{Reaction Rate}
We can express the transition matrix element as,
\begin{eqnarray}
 \langle f|T(t,t_0)|i\rangle &=& i {e^2 Z_X m_2\over 2 \hbar^2 M' V} 
 \sqrt{1\over \omega_1\omega_2 } I'_{\Omega}\sum_n 
	\int_{t_0}^t dt' e^{i(E_f-E_i+ E_{\gamma 1} + E_{\gamma 2}+E_p)t'/\hbar} \nonumber \\
	&\times&  \sum_n I_f I_a I_b {(E_f-E_n)(E_p+E_n-E_i)\over E_p+E_n-E_i+E_{\gamma 1}}
 \label{eq:transitionMat1}
\end{eqnarray} 
where $E_{\gamma 1} = \hbar \omega_1$ and $E_{\gamma 2} = \hbar \omega_2$.
As mentioned earlier, we also need to add another amplitude in 
which the two photons are reversed. However, in most of the kinematic regime
only one of the
two amplitudes dominate and we can neglect the other. This is
because we obtain dominant contribution from the neutron momentum $k_n$, and
hence $K$, close
to the resonant value. Since $\vec K= \vec k_1 + \vec k_{\gamma 1}$, for a 
fixed final state proton momentum, it is clear that we obtain dominant 
contribution from a narrow range of photon momentum $\vec k_{\gamma 1}$ emerging from 
the first vertex. In other words, for any event, given the observed proton
momentum, the photon momentum $\vec k_{\gamma 1}$ is fixed within a 
narrow range, with the corresponding value of $\vec k_{\gamma 2}$ being 
very different. 
Hence the second amplitude in which  
 the photon with 
momentum $\vec k_{\gamma 1}$ emerges from the second vertex will
give negligible contribution and, for brevity, we neglect it. 
The sum over the intermediate neutron states can be computed by converting
it into an integral $\int Vd^3k_2/(2\pi)^3$. 

The transition rate can now be expressed as 
\begin{equation}
    {dP\over dt} = {1\over \Delta T}\int {Vd^3k_p Vd^3k_1 Vd^3k_2\over (2\pi)^9} |\langle f|T(t_0,t)|i\rangle|^2
\end{equation}
where $\Delta T$ is the total time.
The time integral in the transition matrix element is proportional to $\Delta T\delta(E_f-E_i+E_{\gamma 1} + E_{\gamma 2}+E_p)$.   
The reaction rate can now be expressed as
\begin{eqnarray}
	{dP\over dt} &=& {\alpha^2 \over 64\pi^5\hbar^3 c^2} \left({Z_Xm_2\over M'}\right)^2
	V\int {k_p^2 dk_p } d\Omega_p dk_{\gamma 1} k_{\gamma1} d\Omega_{\gamma 1} 
	E_{\gamma2}     |I'_{\Omega}|^2
\nonumber\\
	&\times &\Bigg| {V\over (2\pi)^3}\int dk_nk_n^2 d\Omega_n I_f I'_a I_\Omega I_{b} 
	{(E_f-E_n)(E_n+E_p-E_i)\over E_p+E_n-E_i+E_1}\Bigg|^2
	\label{eq:rate3}
\end{eqnarray}
where $d\Omega_{\gamma 1}$ and $d\Omega_{\gamma 2}$ 
are the measures for the angular 
integrations over the two photon momenta and 
$d\Omega_2$ corresponds to the intermediate state neutron. 
The integration over the proton angular variables is trivial and just gives $4\pi$. 

The reaction rate for the deuteron-nickel system is found to be approximately
100 per sec. This is a fairly large rate and easily observable experimentally.
The photon energy $E_2$ shows a broad peak centered at approximately 2 MeV. As 
mentioned earlier, the energy of the first photon $E_1$ for any event is fixed
by the requirement that $K = |\vec k_1 +\vec k_{\gamma 1}|$ is close to the
resonant value, which in the present case is found to be approximately 1140 in
atomic units.

In our calculation we have focussed on the contribution from a single 
isolated resonance. It is important to determine whether the contribution
from other resonances may change our result significantly. We have tested
this possibility by assuming a resonance at a higher value of $k_n$ with
the value of $K$ fixed to approximately 1140, applicable for the resonance
under consideration. We find that the amplitude falls exponentially if 
the value of $k_n$ is widely different from $K$ and hence the contribution
from other resonances is expected to be negligible.

\section{Conclusions}
In this paper we have  
computed the rate for the nuclear reaction given in Eq. \ref{eq:HXfusion} at second order in time dependent perturbation theory. The initial state consists of a deuteron and a heavy nucleus ${}^A$X at very low energies, of order eV or less. The reaction proceeds by formation of an intermediate state which is composed of a free proton and neutron. Despite 
 the low energy of the initial state, such a state can exist for a small time 
 interval. Due to the small time interval, this leads to a suppression factor in the rate of this process. However this suppression is not as prohibitive as the very strong suppression arising due to Coulomb barrier. Hence the rate of this process may be substantially higher in comparison to expectations based on the standard Coulomb barrier.
 The neutron gets captured
by the heavy nucleus which is assumed to happen with emission of a photon, 
although other reactions may also be possible. Overall the final state consist of a proton, two photons and the nucleus ${}^{A+1}$X.  

We have computed the rate assuming that the initial state deuteron is bound to the heavy nucleus with the formation of a molecule,
with the wave function given by Eq. \ref{eq:psicm}.  
Furthermore we have considered, as an example, the heavy nucleus to be
${}^{58}$Ni which absorbs a neutron into the $2p_{3/2}$ level.
We find that the rate for such a process is, in general, 
very small. However, in the presence of a resonance the rate may be substantial
and observable \cite{ramkumar2024low}. We find that there exists an 
isolated resonance in the neutron-nickel system at energy of 7 keV, which
has a width of approximately 7 eV. Assuming that this resonance dominates
the amplitude, we find that the rate for the process is approximately 100
per second, which is easily observable. This rate turns out to be much 
larger than that found for the weak interaction process considered in 
\cite{ramkumar2024low}. The process leads to emission of two photons
and a proton, with one of the photons showing a broad peak at approximately
2 MeV. Furthermore, the total momentum of the second photon and proton should
show a peak at 1140 in atomic units, i.e. close to the resonant value $k_R$.

We have pointed out that besides the process considered, several other
transitions are possible in which only one or no     
photon may be emitted. Alternatively, the transition may not be to the
ground state of ${}^{59}$Ni. In this case, 
the two primary photons would have lower
total energy and will be followed by a 
cascade of lower energy photons. These lower energy
photons have a much higher probability of being captured by the medium and
hence will lead to excess heat in the medium. Excess heat will also be
generated by the nuclear recoil, which will lead to transfer of energy to the medium. Hence we also expect to see lower energy primary photons along with
a cascade of lower energy photons and excess heat in the medium. 
A detailed study of this process is postponed to future research.

 \bigskip
 \noindent
% {\bf Acknowledgements:} We are very grateful to 

\bibliographystyle{ieeetr}
\bibliography{nuclear}
\end{document}